\documentclass[aps,showpacs,prl,twocolumn]{revtex4-1}  

\usepackage{amsmath}
\usepackage{mathrsfs}
\usepackage{amssymb}
\usepackage{graphicx}
\usepackage{bm}
\usepackage{color}
\usepackage{times}
\usepackage{hyperref}

\definecolor{clr}{rgb}{0,0.6,0.6}

\begin{document}

\title{Cooperative atom-light interaction in a blockaded Rydberg ensemble}

\author{J. D. Pritchard}
\email{j.d.pritchard@durham.ac.uk}
\author{D. Maxwell}
\author{A. Gauguet}
\author{K. J. Weatherill}
\author{M. P. A. Jones}
\author{C. S. Adams}
\email{c.s.adams@durham.ac.uk}
\affiliation{Department of Physics, Durham University, Rochester Building, South Road, Durham DH1 3LE, UK}

\date{\today}

\begin{abstract}
By coupling a probe transition to a Rydberg state using electro-magnetically induced transparency (EIT) we map the strong dipole-dipole interactions onto an optical field. We characterize the resulting cooperative optical non-linearity as a function of probe strength and density. We show that the effect of dipole blockade cannot be described using a mean-field but requires an $N$-atom cooperative model. Good quantitative agreement is obtained for three atoms per blockade with the $n=60$ Rydberg state. We place an upper-limit on the dephasing rate of the blockade spheres of $<110$~kHz.
\end{abstract}


\maketitle
In atomic ensembles one can distinguish between collective and cooperative phenomena. In a collective process the response of the system is given by a sum of the local response of each atom. In a cooperative process however, the interactions between atoms are sufficiently strong that one can no longer separate the response of each atom from that of others \cite{mandel08}. An example of cooperativity is superradiance, where the constructive interference between individual dipoles leads to a modification of the rate and angular distribution of spontaneous emission \cite{dicke54,gross82}. Similarly, dipole-dipole interactions may also modify the absorptive and dispersive properties of a medium \cite{hehlen94}. For a two level system, the dipole-dipole induced modification of the optical response becomes significant when the interatomic spacing $R$ is less than $\lambda/2\pi$, where $\lambda$ is the transition wavelength. For optical transitions the condition $R<\lambda/2\pi$ leads to other broadening mechanisms and consequently cooperative effects due to dipole-dipole interaction have only been observed in an up conversion process \cite{hehlen94}. In Rydberg atoms the condition $R<\lambda/2\pi$ is easily fulfilled as the strongest dipole allowed transitions are in the microwave domain. In this case, the condition that the dipole-dipole interaction is larger than the linewidth gives rise to the blockade effect \cite{lukin01} that has been used to realize strongly correlated quantum gases \cite{heidemann07}, entanglement \cite{wilk10} and quantum gates \cite{isenhower10}. To control the optical response one can map the properties of the Rydberg states onto a strong optical transition using a dark state \cite{mohapatra07,mohapatra08}. Similarly, Rydberg dressing of the ground state has also been proposed as a technique to realize novel quantum phases \cite{pohl10,pupillo10}. 

In this paper we demonstrate a cooperative atom-light interaction due to dipole blockade of the Rydberg state in an ultra-cold atomic ensemble. The effect of strong interactions between Rydberg pairs is mapped onto an optical transition using electromagnetically induced transparency (EIT) \cite{boller91,mohapatra07} resulting in an optical non-linearity. As EIT probes a dark state consisting of a superposition of the ground and Rydberg states, the lineshape is sensitive to the coherence of the blockaded ensemble. We observe no additional dephasing of the dark state as the strength of the optical field, and hence the blockade effect, is increased. This result is inconsistent with a theoretical description of individual atoms coupled to a mean--field, where one finds that the modification of the dark state due to interactions is always accompanied by dephasing. We show that a cooperative model describing the dynamics of the full $N$--atom system provides excellent quantitative agreement for $N=3$. The demonstrated cooperative optical non-linearity is significant in the context of  photonic quantum gates \cite{friedler05} and single photon sources \cite{saffman02,pedersen09}. 

\begin{figure}[t]
\includegraphics{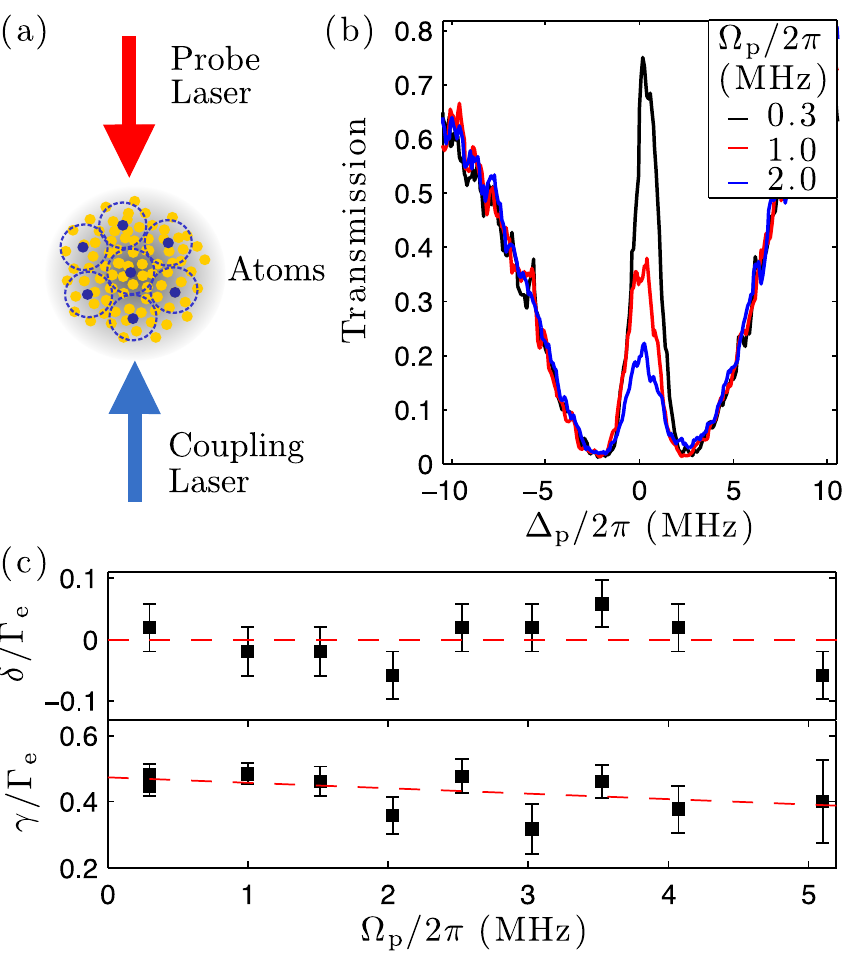}
\caption{(Color online). (a) Schematic of experiment. EIT spectroscopy is performed on an ultra-cold $^{87}$Rb atom cloud. (b) Suppression of transparency on resonance for coupling to 60$S_{1/2}$ for increasing probe Rabi frequency $\Omega_\mathrm{p}$ at a density $\rho=1.2\pm0.1\times10^{10}$~cm$^{-3}$. (c) Detuning, $\delta$, and width, $\gamma$, of EIT resonance as  a function of $\Omega_\mathrm{p}$ in units of the excited state width $\Gamma_e = 2\pi\times6$~MHz, showing no dephasing and no resonance shift.}
\label{fig1}
\end{figure}

Our experiments are performed on a laser cooled $^{87}$Rb atom cloud using the experimental setup described in \cite{weatherill08} and shown schematically in Fig.~\ref{fig1}(a). Atoms are loaded into a magneto-optical trap for 1~s, reaching a temperature of 20~$\mu$K. Atoms are then prepared in the 5s $^2$S$_{1/2}~\vert F=2,m_F=2\rangle$ state ($\vert g \rangle$) by  optical pumping. By varying the optical pumping duration, the fraction of atoms in $F=2$ and hence the density in state $\vert g \rangle$ can be controlled without changing the cloud size. EIT spectroscopy is performed using counter--propagating probe and coupling lasers focused to 1/$e^2$ radii of 12~$\mu$m and 66~$\mu$m respectively. The coupling laser is frequency stabilized to the 5p $^2$P$_{3/2}~F'=3\rightarrow$ ns~$^2$S$_{1/2}$ transition using an EIT locking scheme \cite{abel09}. The probe laser drives the 5s $^2$S$_{1/2}~F=2$ $\rightarrow$ 5p $^2$P$_{3/2}~F'=3$ transition and is scanned across the resonance from $\Delta_\mathrm{p}/2\pi = -20\rightarrow+20$~MHz in 500~$\mu$s. The probe and coupling lasers are circularly polarized to drive $\sigma^+$--$\sigma^-$ transitions, maximizing the transition amplitude to the Rydberg state. Probe powers in the range 1~pW to 5~nW were used to explore the optical non-linearity. The transmission is recorded using a single photon counting module. To give a constant signal to noise ratio, the probe beam is attenuated after the atoms to give a fixed count rate of 1.5~Mc/s at the detector for each power. For each power, the experiment is repeated 100 times to build up a transmission histogram. The width of the atom cloud along the probe axis, $\ell$, was measured by fluorescence imaging of the cloud after preparation in $F=2$, giving $\ell = 1.4\pm0.1$~mm. To measure the density of atoms in the probe beam, transmission data is recorded with the coupling laser off. The spectra are fitted using the analytic absorption profile for a two-level atom \cite{loudon08} to extract both the density and the effective Rabi frequency due to the saturation of the transition and Gaussian intensity profile. We obtain a peak density of  $\rho = 1.2\pm0.1\times10^{10}~\mathrm{cm}^{-3}$ giving around 7000 atoms in the interaction region.

Figure~\ref{fig1}(b) shows EIT spectra as a function of probe Rabi frequency, $\Omega_\mathrm{p}$ for coupling to the 60$S_{1/2}$ Rydberg state. As the Rabi frequency is increased, there is a dramatic suppression of transmission on the two-photon resonance from 75~\% to 20~\%. This intensity dependent transmission on resonance gives an optical non-linearity. The effect saturates at $\Omega_\mathrm{p}/2\pi = 5$~MHz. From the lineshape the detuning ($\delta$) and the FWHM ($\gamma$) of the two-photon resonance are extracted as a function of $\Omega_\mathrm{p}$, see Fig.~\ref{fig1}(c). This shows there is neither a broadening nor a detuning of the EIT feature to accompany the suppression. The absence of a shift or broadening is significant, as it rules out inhomogeneous broadening mechanisms such as Stark-shift due to ions or van der Waals dephasing of the Rydberg state \cite{gross82}.

\begin{figure}[t]
\includegraphics{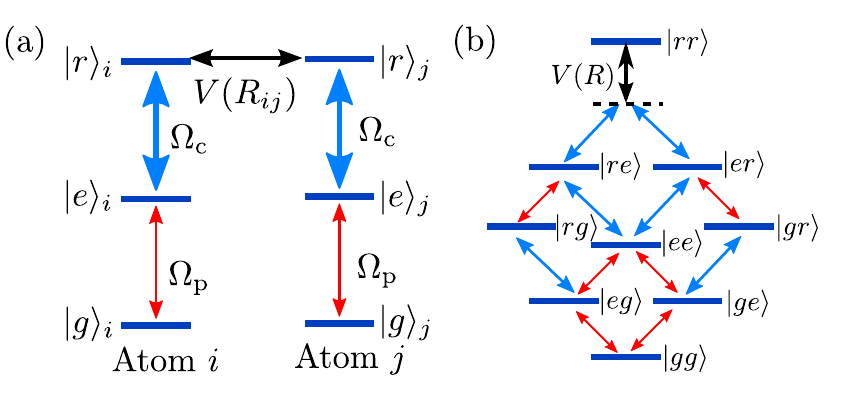}
\caption{(Color online). Schematic of exact $N$-atom interaction model where all pair-wise dipole-dipole interactions between atoms $i$ and $j$ are included. (b) Energy levels for $N$=2 atom model. The dipole-dipole interaction term $V(R)$ acts to detune the doubly excited state off resonance, leading to blockade.}
\label{fig2}
\end{figure}

The observation of suppression without shift or broadening is explained by considering the dipole-dipole interactions between Rydberg atoms \cite{mueller08}. For $S$ states the long-range dipole-dipole potentials have the form $V(R)=-C_6/R^6$, where $R$ is the interatomic separation and $C_6<0$ \cite{amthor07a}. We consider a many-body system of $N$-atoms, where each pair is coupled via the dipole-dipole interaction, shown schematically in Fig.~\ref{fig2}(a). The resulting $N$-atom Hamiltonian $\hat{\mathscr{H}}_N$ is given by:
\begin{equation}
\hat{\mathscr{H}}_N = \displaystyle\sum_i^N \hat{\mathscr{H}}^{(i)}  - C_6\displaystyle\sum_{j<i} \frac{\hat{P}^{(i)}_{rr}\hat{P}^{(j)}_{rr}}{{R}_{jj}^6}
\label{eq1}
\end{equation}
where $\hat{\mathscr{H}}^{(i)}$ is the single atom Hamiltonian describing the coupling to the probe and coupling field acting on atom $i$ and $\hat{P}^{(i)}_{rr}=\vert r \rangle_{ii}\langle r \vert$ is the projector onto the Rydberg state of the $i^\mathrm{th}$ atom. For a given set of parameters the temporal evolution is calculated from the master equation $\dot{\hat{\sigma}} = i/\hbar [\hat{\sigma},\hat{\mathscr{H}}_N]-\hat{\gamma}$ where $\hat{\sigma}$ is the $3^N\times3^N$  density matrix for the $N$ atom system and ${\gamma}$ is the decoherence matrix, including the dephasing due to finite laser linewidth. The susceptibility, and the transmission, is calculated by summing over all the coherence terms in the density matrix between states that are coupled by the probe laser.

The effect of dipole-dipole interactions on the two-photon resonance can be seen from considering the simplest case of $N=2$, shown in Fig.~\ref{fig2}(b). In the absence of interactions $(V(R)=0)$ the system evolves into the following eigenstate on the two-photon resonance:
\begin{equation}
\vert D\rangle = \frac{\Omega_\mathrm{c}^2 \vert gg\rangle  - \sqrt{2}\Omega_\mathrm{p}\Omega_\mathrm{c} \vert gr\rangle^+ + \Omega_\mathrm{p}^2 \vert rr\rangle}{\Omega_\mathrm{p}^2+\Omega_\mathrm{c}^2},
\label{eq2}
\end{equation}
where $\vert gr \rangle^+ = (\vert gr\rangle+\vert rg\rangle)/\sqrt{2}$, $\Omega_\mathrm{c}$ is the coupling Rabi frequency and the relative phase between the lasers has been neglected. This is a dark state as it is not coupled to the probe laser, leading to the observed transparency. 

Dipole-dipole interactions between the Rydberg states modify this dark state by suppressing excitation of $\vert rr \rangle$ when $V(R)>\Omega_\mathrm{c}$ due to blockade. The resulting state can no longer be written as a product state and involves the components $\vert gg \rangle, \vert ge\rangle^+, \vert gr\rangle^+$ and $\vert er\rangle^+$. Due to the rapid decay of state $\vert e \rangle$ this state differs from the eigenstate found in \cite{moller08}. The physical interpretation of this result is that the system is in a superposition where only one atom can be excited to the Rydberg state and thereby contribute to the EIT dark state, whilst the other acts as a two-level atom which couples resonantly to the probe field. This leads to a suppression of transmission on resonance. Actually the blockade forms a collective state where the single Rydberg excitation is shared between the two atoms. As more atoms are included in the blockade sphere, the single excitation is now shared across a larger number of atoms. This leads to enhanced suppression as the fraction of atoms contributing to EIT is reduced. 

\begin{figure}[t]
\includegraphics{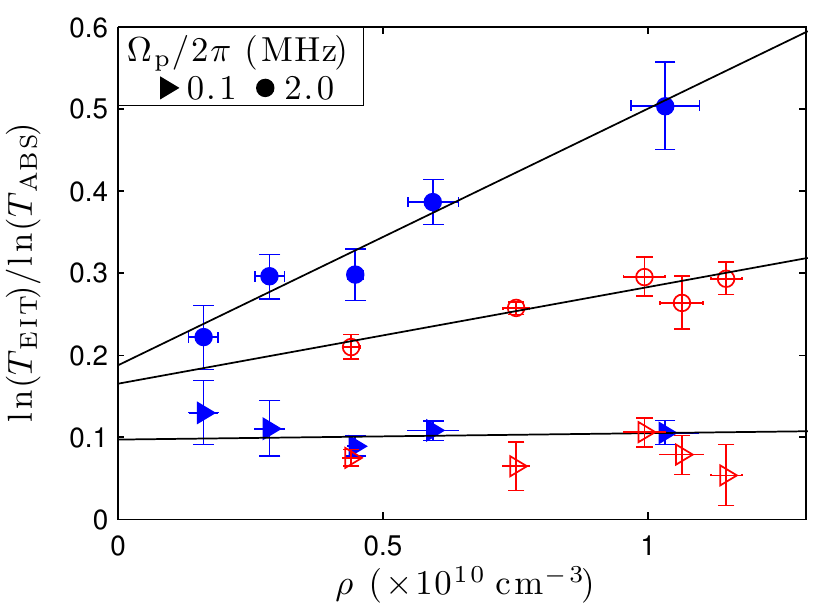}
\caption{(Color online). Optical depth as a function of density for 60$S_{1/2}$ (\textcolor{blue}{$\blacktriangleright$}, \textcolor{blue}{$\bullet$}) and 54$S_{1/2}$ (\textcolor{red}{$\vartriangleright$}, \textcolor{red}{$\circ$}) for weak and strong $\Omega_\mathrm{p}$ respectively, scaled by probe only optical depth to remove trivial linear scaling. Strong probe data reveals second order density scaling consistent with a cooperative optical non-linearity. Comparison of the gradients for $60S_{1/2}$ and $54S_{1/2}$ gives a ratio of $2.6\pm0.7$.}
\label{fig3}
\end{figure}

We demonstrate the scaling of the non-linearity with number of atoms in each blockade sphere, $N$, by changing both the density and the principal quantum number, $n$. In an ordinary non-linear medium, the optical response scales linearly with density as each optical dipole is independent. For a cooperative effect due to dipolar interactions, the optical response is now proportional to the number of atoms in a blockade sphere, leading to a non-linear scaling with density. Figure~\ref{fig3} shows the optical depth on resonance as a function of density for 60$S_{1/2}$ and 54$S_{1/2}$ for both strong and weak probe powers. To remove the trivial linear response with density the optical depth is scaled by the value with coupling laser off. In the weak probe regime, the resonant eigenstate is equivalent to $\vert D\rangle$ and no non-linear scaling is observed. For the strong probe data, there is a second-order scaling with density, consistent with increased suppression due to an increase in the number of atoms in each blockade sphere. Taking into account the scaling of $C_6$ and $\Omega_\mathrm{c}$ with $n$, the number of atoms in a blockade sphere should scale as $N\propto n^{6.25}$. The ratio of $N$ for 60$S$ to 54$S$ is 2.0, whilst the ratio of the linear fit gradients gives a ratio of $2.6\pm0.7$. This is consistent with the blockade scaling, showing the tuneability of the cooperative optical non-linearity with principal quantum number.

To reproduce the data shown in Fig.~\ref{fig1} it is necessary to solve the many-body Hamiltonian for the average number of atoms in each blockade sphere. For the $60S_{1/2}$ state, the van der Waals coefficient is  $C_6 = -140$~GHz~$\mu$m$^6$~\cite{singer05}, which for $\Omega_\mathrm{c}/2\pi=4.6~$MHz gives a blockade radius $r_\mathrm{b}=\sqrt[6]{C_6/\Omega_\mathrm{c}}=5.6~\mu$m. For a density of 1.2$\times10^{10}$~cm$^{-3}$ this gives an average of $\bar{N}$= 9 atoms in each blockade sphere. As a full solution of Eq.~\ref{eq1} for large $N$ is demanding, previous work on $N$-atom blockaded systems have focussed on a mean-field description \cite{tong04, weimer08, chotia08, raitzsch09}. However, a mean-field approach is insufficient to reproduce the experimental observations shown in Fig.~\ref{fig1} as the average interaction experienced by each atom by its neighbors contributes to blue detuning of the Rydberg state. This leads to both broadening and a shift of the EIT lineshape when summing over each atom \cite{schempp10}. No dephasing is observed in our experiment because the blockade suppresses multiple Rydberg excitations and consequently van der Waals dephasing is also suppressed. 

\begin{figure}[t]
\includegraphics{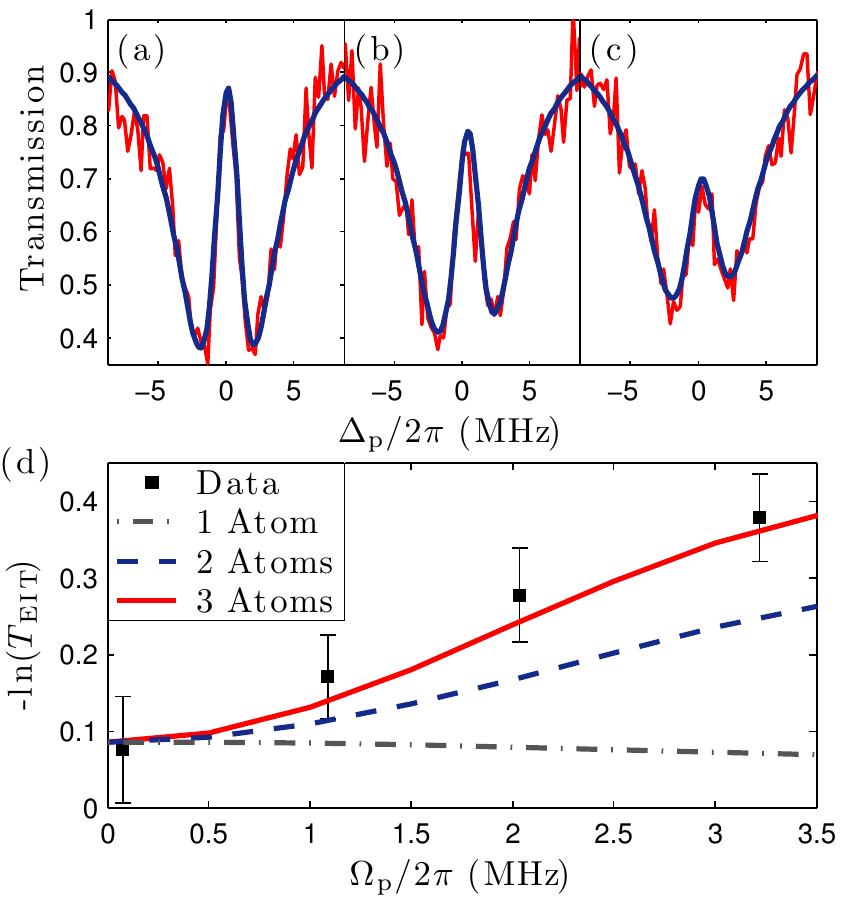}
\caption{(Color online). Comparison of $N$-atom model to transmission data at a density of $\rho=0.35\pm0.03\times10^{10}~$cm$^{-3}$ with $\bar{N}=3$. Traces (a) -- (c) show spectra recorded at $\Omega_\mathrm{p}/2\pi = 1.0,~2.0,~3.2$~MHz respectively with the three-atom model plotted on top (thick line). (d) Optical depth on resonance as a function of $\Omega_\mathrm{p}$ compared to model for $N$=1, 2 and 3 atoms. Good qualitative agreement is obtained for $N$=3 to both resonant transmission and lineshape. All curves calculated using $\Omega_\mathrm{c}/2\pi = 3.8~$MHz and $V(R)/2\pi$=15~MHz with linewidths of the probe laser and two photon transition of 0.2 and 0.11~MHz respectively.} 
\label{fig4}
\end{figure}

To obtain quantitative agreement with the full $N$-atom model we reduce the density to $0.35\pm0.03\times10^{10}$~cm$^{-3}$ and $\Omega_\mathrm{c}/2\pi$ to 3.8~MHz to give an average of $\bar{N}=3$ atoms per blockade sphere. Transmission data for $\Omega_\mathrm{p}/2\pi=$~0.1~to 3.2~MHz is shown in Fig.~\ref{fig4} (a)--(c), again showing suppression of the resonant transmission but by a smaller amount as the number of atoms per blockade is reduced. Also shown on the data is the transmission calculated using the many-body model for $N=3$. Model parameters are determined by matching the single atom model to the weak probe transmission for $\Omega_\mathrm{p}/2\pi = 0.1$~MHz, yielding the coupling Rabi frequency and linewidths of the probe laser and two-photon resonance of 3.8, 0.2 and 0.1~$(\times2\pi)~$MHz respectively, consistent with experimental parameters. Transmission traces are then calculated changing only $\Omega_\mathrm{p}$. There is only one free parameter in the model, which is the interaction strength, $V$. For the model traces in Fig.~\ref{fig4} calculations were performed using $V/2\pi = 15$~MHz however the model is insensitive to $V$ once the blockade condition ($V>\Omega_\mathrm{c}$) is met. The model shows very good quantitative agreement with the data, reproducing both the suppression on resonance and the EIT lineshape. Fig.~\ref{fig4}(d) shows the optical-non-linearity by plotting the optical depth on the two-photon resonance as a function of $\Omega_\mathrm{p}$ compared to the model for $N$=1, 2 and 3. The small decrease with $\Omega_\mathrm{p}$ observed for $N$=1 is due to the system being driven faster relative to the dephasing caused by finite laser linewidth. Comparing this to the curves for $N$=2 and 3 shows that the transmission is strongly modified compared to the case for a single atom, with a good agreement to the $\bar{N}$=3 data as expected.

The results of Fig.~\ref{fig1}(c) and Fig.~\ref{fig4} show that EIT is sensitive to the coherence of the blockaded ensemble, as only by considering the coherence terms of the complete many-body system is it possible to reproduce the suppression of transmission without introducing broadening of the resonance. In the weak probe limit the width of the EIT resonance is dominated by the relative linewidth of the probe and coupling laser. Fitting the weak probe data gives a linewidth of $110\pm50$~kHz. As no additional broadening is observed at increased probe power, this places an upper limit on the dephasing rate of each blockade sphere of $<$110~kHz for 60$S_{1/2}$.

In summary, we explore the effects of dipole-dipole interactions between Rydberg atoms on light propagation. By mapping the strong interactions onto an optical field, a novel cooperative optical non-linearity is observed. This differs from other cooperative effects such as superradiance, where the cooperativity is mediated by the optical field. Instead, the cooperativity arises from long-range dipole-dipole interactions between Rydberg states that is observed as a back-action on the probe field. As the optical response of each atom is significantly modified by its proximity to neighboring atoms, the dynamics can only be described by treating all $N$ atoms in each blockade sphere. We show excellent quantitative agreement between experiment and theory for three atoms per blockade, and verify the expected density scaling with the strength of the van der Waals interaction. By probing the coherence of the blockaded system we place an upper limit on the dephasing rate of each blockade sphere of $<$110~kHz, enabling potential applications in quantum optics. Future work will focus on using the observed non-linearity to develop single photon sources by working with a single blockaded ensemble.

\begin{acknowledgments}
We thank M. M\"{u}ller and I. Lesanovsky for help with elucidating the theoretical origin of the suppression effect. We are also grateful R. M. Potvliege, I. G. Hughes and T. Pfau for stimulating discussions, and acknowledge support from the UK EPSRC.
\end{acknowledgments}


%

\end{document}